\documentclass{article}
\usepackage[colorlinks,linkcolor=blue]{hyperref}
\usepackage{spconf,amsmath,graphicx}
\usepackage{multirow}
\usepackage{amsfonts}
 \usepackage[moderate]{savetrees}
\usepackage{booktabs}  %
\usepackage{hyperref}

\usepackage{enumitem}
\title{VoxBlink: A Large Scale Speaker Verification Dataset on Camera}
\name{Yuke Lin$^{1,2}$,Xiaoyi Qin$^{1,2}$, Guoqing Zhao$^{3}$,Ming Cheng$^{1,2}$, 
Ning Jiang $^{3}$,Haiying Wu$^{3}$,
Ming Li$^{1,2}$\thanks{Corresponding Author: Ming Li. \\This research is funded in part by the National Natural Science Foundation of China (62171207), Science and Technology Program of Suzhou City(SYC2022051) and MaShang Consumer Finance Co.Ltd. Many thanks for the computational resource provided by the Advanced Computing East China Sub-Center. }}
\address{
$^1$School of Computer Science, Wuhan University, Wuhan, China \\
$^2$Suzhou Municipal Key Laboratory of Multimodal Intelligent Systems, Duke Kunshan University, Kunshan, China \\
$^3$Mashang Consumer Finance Co., Ltd 
}
\begin{document}
%
\maketitle
\begin{abstract}
In this paper, we introduce a large-scale and high-quality audio-visual speaker verification dataset, named \textbf{VoxBlink}. We propose an innovative and robust automatic audio-visual data mining pipeline to curate this dataset, which contains 1.45M utterances from 38K speakers. Due to the inherent nature of automated data collection, introducing noisy data is inevitable. Therefore, we also utilize a multi-modal purification step to generate a cleaner version of the VoxBlink, named VoxBlink-clean, comprising 18K identities and 1.02M utterances. In contrast to the VoxCeleb, the VoxBlink sources from short videos of ordinary users, and the covered scenarios can better align with real-life situations. To our best knowledge, the VoxBlink dataset is one of the largest publicly available speaker verification datasets. Leveraging the VoxCeleb and VoxBlink-clean datasets together, we employ diverse speaker verification models with multiple architectural backbones to conduct comprehensive evaluations on the VoxCeleb test sets. Experimental results indicate a substantial enhancement in performance—ranging from 12\% to 30\% relatively—across various backbone architectures upon incorporating the VoxBlink-clean into the training process. The details of the dataset can be found on \href{http://voxblink.github.io}{Site}.
\end{abstract}

\section{Introduction}
\label{sec:intro}
Automatic Speaker Verification (ASV) in wild scenarios has achieved remarkable success consisting of the evolution of backbone architecture \cite{desplanques20_interspeech,cai18_odyssey,zhang22h_interspeech}, the introduction of diverse loss functions \cite{8953658,Liu_2017_CVPR}, and the availability of large-scale corpora \cite{Nagrani17,chung18b_interspeech}. Even though growing efforts have been devoted to refining networks and training strategies \cite{9414600,10096659}, the academic community still faces constraints by the limited scale and diversity of available datasets. 

\begin{figure}
  \includegraphics[width=0.4\textwidth]{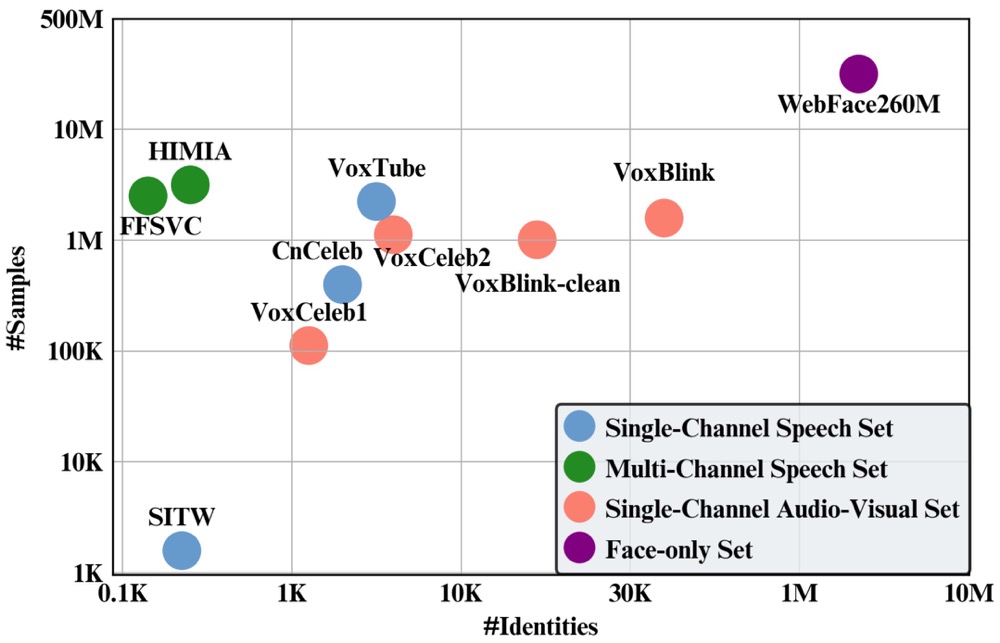}
  \centering
  \caption{{
    Comparisons of \# identities and \# samples for our VoxBlink data
and public ASV training set (also a face dataset as a contrast). The x and y-axis have \textbf{non-uniform} scales. }}
  \label{fig::dataset}
\end{figure}

In the computer vision field, millions of images establish a robust foundation for face recognition. Regrettably, in the field of speaker recognition, the availability of publicly accessible datasets with both millions of utterances and tens of thousands of speakers in the wild remains noticeably limited. As is shown in Fig \ref{fig::dataset}, many contributions have been made to enrich ASV datasets \cite{Nagrani17,chung18b_interspeech,9054017,qin20222022,9054423,mclaren16_interspeech,7178964,yakovlev23_interspeech}. Some datasets \cite{qin20222022,9054423} mainly comprise multi-channel far-field speeches, which includes limited number of speakers. While others \cite{9054017,mclaren16_interspeech,7178964} fall short of their limited styles, languages and scales. Among these endeavors, the VoxCeleb \cite{Nagrani17,chung18b_interspeech} stands out as the most successful as it contains over one million utterances from thousands of speakers.
Nonetheless, compared to face recognition, the quantity remains relatively small. Recently, the VoxTube \cite{yakovlev23_interspeech} dataset releases over 4M utterances for 5,040 speakers, making it one of the largest open-source speaker recognition datasets to date. However, due to its reliance solely on audio information for clustering, the accuracy of its derived labels may not be very convincing. Meanwhile, hard samples can be easily discarded during the filtration. 

Therefore, we introduce a new large-scale audio-visual dataset for speaker verification, VoxBlink. All VoxBlink data is captured automatically from users who upload \textit{short} videos on the YouTube platform, which contain over \textbf{1.4M utterances} from over \textbf{38K speakers}. In order to further purify the data without filtering out difficult samples, we use a multi-modal validation approach that results in a purified version of the VoxBlink (VoxBlink-clean), which contains 18k individuals and 1.02M utterances. All raw data is processed by multi-stage audio-visual models (including but not limited to face and lip detection, face verification, active speaker detection and overlap detection). Furthermore, by implementing audio-visual models for data mining, our data automatic pipeline exhibits greater resilience and promise in data size. Since data from the VoxBlink are mainly collected from the ``wild'', it also inherently exhibits diverse-age, diverse-lingual, diverse-style and diverse-device attributes. As the VoxBlink is an audio-visual dataset, it can be used in various other applications, such as speech separation\cite{10023284,10094306}, multi-modal verification\cite{10095883,sadjadi20202019}, and speaker diarization\cite{10095802,wang2023multimodal}, among others. 

In addition, we also incorporate the VoxBlink-clean dataset in training various models with different backbones. When introducing the VoxBlink-clean, all models exhibit a performance boost ranging from 12\% to 30\% in terms of relative EER reduction on the VoxCeleb1-O test set. Our primary contributions can be summarized as follows:
1) We propose a more scalable and robust pipeline for mining speaker verification data; 2) We collect a large-scale audio-visual speaker verification dataset VoxBlink and its purified version VoxBlink-clean; 3) We achieve significant performance improvements under different backbones by integrating the VoxBlink-clean dataset into training process. 


\section{DATA MINING}
\label{sec:VoxBlink}
\subsection{Data Description}
\begin{figure}[t]
  \includegraphics[clip, trim=0cm 0cm 0cm 0cm,width=0.4\textwidth]{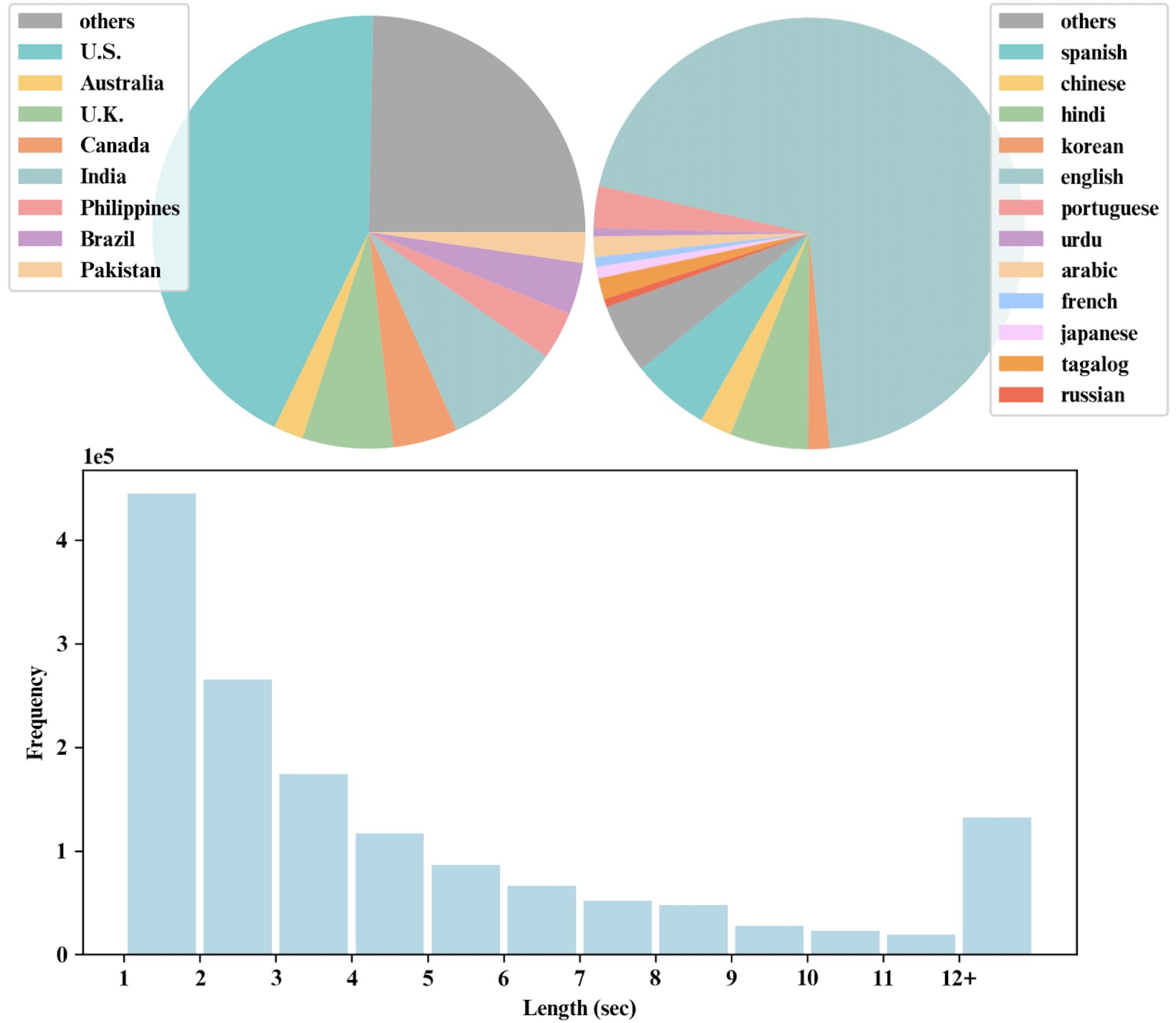}
  \centering
  \caption{{
    \textbf{Top:} the distribution of geographic locations (\textit{left}) and languages (\textit{right}) of speakers.
    \textbf{Bottom}: The distribution of utterance lengths in the dataset.
     }}
  \label{fig::statistics}
\end{figure}

The VoxBlink contains 1,455,190 utterances from 38,065 channels on YouTube, while its purified version VoxBlink-clean comprises 1,028,095 utterances from 18,381 speakers. All speech/video segments are extracted from short videos uploaded by ordinary YouTube users, encompassing various contexts, including podcasts, music lives, speeches, live streaming highlights, etc. Indoor reverberation, non-verbal voice, background music and other acoustic conditions have increased the complexity and diversity of the data. Most of the segments are recorded on mobile devices, with recording environments spanning indoors, outdoors, and a variety of complex scenarios. Other statistic information can be found in Table \ref{tab::statistics} and Fig \ref{fig::statistics} shows a visualization of the statistics. The majority of speakers within the VoxBlink dataset are female, and its purified version is relatively gender-balanced. The dataset is multi-lingual yet English-dominant, with participants ranging from over 130 different regions worldwide. 

\begin{table}[ht]\centering \footnotesize
\caption{\label{tab::statistics} {\it Statistics for the VoxBlink dataset. The last two rows describe the \textbf{time-varying} characteristics across videos recorded by the same speaker.}}
\begin{tabular}{@{}lcc@{}}
\toprule
\textbf{Dataset} & \textbf{VoxBlink} & \textbf{VoxBlink-clean} \\ 
\midrule
\# of SPKs  & 38,065 & 18,381 \\
\# of male-SPKs  & 15,013 & 8,124 \\
\# of videos & 372,084 & 241,170 \\
\# of hours & 2,135 & 1,670 \\ 
\# of utterances & 1,455,190 & 1,028,095 \\
Avg \# of videos per SPK & 9.77 & 13.12 \\
Avg \# of utterances per SPK & 38.23 & 55.93 \\
Avg \# of duration per utterance (s) & 5.28 & 4.87 \\
Avg \# of video recording intervals (days) & 39.72 & 34.55 \\
Avg \# of video recording span (days) & 440.07 & 441.85 \\
\bottomrule
\end{tabular}
\end{table}

\subsection{Collection Pipeline}
As depicted in Fig \ref{fig::pipeline}, we employ an automatic multi-modal data-mining pipeline to construct the VoxBlink database from YouTube. The novelty of our approach lies in utilizing user avatars for frame-by-frame face verification and lip motions for active speaker detection. Additionally, with the help of other auxiliary tools, we can extract speech/video segments specifically pertaining to the target user. For clarity, we summarize the processes as follows:

\begin{figure}[t]
  \includegraphics[width=0.5\textwidth]{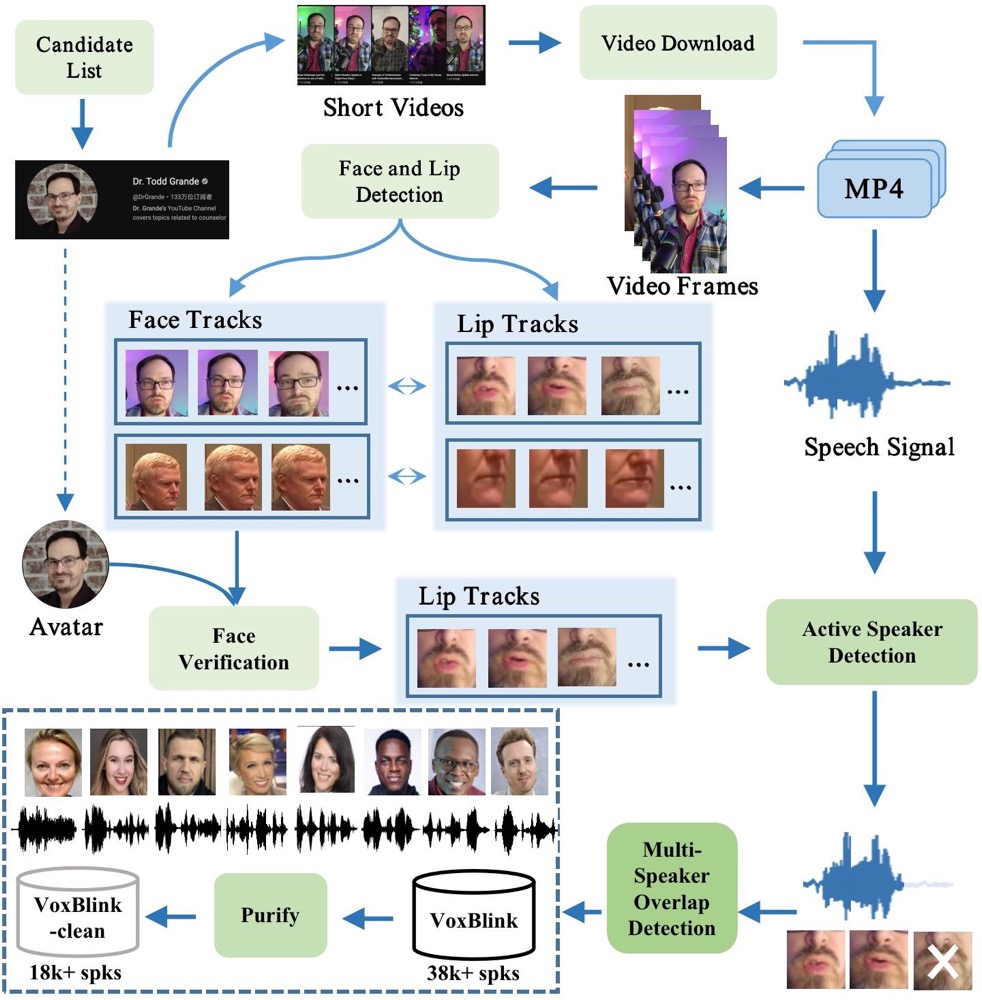}
  \centering
  \caption{{
    The automatic pipeline for the VoxBlink dataset.
    }}
  \label{fig::pipeline}
\end{figure}



\noindent \textbf{Step I: Candidate Collection. } We start by compiling a list containing over 2,000 commonly used names as well as some professions and themes for data diversity. Having observed that users are more likely to appear in short videos, we opt to retain users with a single-face avatar who have uploaded short videos. Over 1M videos from 61,038 users with avatars are downloaded after duplicate removal in the YouTube retrieval. Since the data source of the VoxBlink consists of only short videos, there should be no overlapping with the VoxCeleb and VoxMovies\cite{Brown20b} datasets.

\noindent \textbf{Step II: Face and lip tracking.} Using the Retina Face\cite{9157330} model, we detect both face and lip movements, producing corresponding video and lip tracks. By setting a threshold for the minimum Intersection Over Union (IOU) value between two consecutive detections, we ensure that each track contains only one face or lip sequence.

\noindent \textbf{Step III: Face verification.}  After face and lip tracking, we utilize the ResNet-IRSE50 \cite{8953658} model to extract face embeddings for each speaker frame by frame along each track. Meanwhile, the face embedding of the avatar has been extracted as template embedding using flip augmentation, which promotes template robustness. Then, cosine similarities are calculated along the track, and the track-level average score is calculated to discard non-target tracks.

\noindent \textbf{Step IV: Active Speaker Detection.} To refine the track and eliminate silent or out-of-sync segments, we utilize a Seq2Seq audio-visual speaker diarisation model \cite{10095802}. This model leverages lip motions and audio cues to identify instances of active speech within the track. This approach not only facilitates the removal of silent or voice-over sections but also effectively excludes out-of-sync fragments.

\noindent \textbf{Step V: Multi-Speaker Overlap Detection.} In order to mitigate the potential disruption caused by overlapping speech data and enhance the quality of speech segments, we employ a conformer-based Overlapping Speech Detection (OSD) toolkit \cite{cheng2023dku}. Furthermore, we discard utterances shorter than one second in duration.

\noindent \textbf{Step VI: Meta Information Collection.} 
Due to platform constraints, only the geographical locations of approximately 21,000 speakers are recorded. Given the challenges in obtaining gender labels, we binary-classify the speakers' genders using audio-visual data. We also infer the utterance-level language labels using the Whisper's\cite{radford2023robust} base model and obtain speaker-level labels through a voting mechanism. Other meta-informations about the video, including each video's release time, category, and tags, are collected for other potential applications. All collection meta will be released together with data.

Finally, The audio-visual data obtained through the pipeline constitutes the VoxBlink, whose name is inspired by the characteristics of short videos. The threshold within the collection pipeline is intentionally relaxed to facilitate the accumulation of a larger volume of data.

\subsection{Multi-modal Purify}
Due to automated data collection and relatively relaxed threshold settings, the introduction of noisy data is inevitably unavoidable. Upon manual inspection, we have found that some recordings are merely lip-syncing, indicating the need to purify our data further. Therefore, we purify the VoxBlink dataset utilizing the following metrics:
\begin{itemize}
\setlength\itemsep{0em}
\setlength\topsep{0em}
\setlength\partopsep{0em}
\setlength\parsep{0em}
    \item The average score of within-speaker speech embedding cosine similarities derived by ResNet34 \cite{cai18_odyssey}.
    \item The average score of within-speaker face embedding cosine similarities derived by \cite{8953658}.
    \item The average music-speech discrimination score of a speaker by \cite{lee2017sample}.
\end{itemize}
We retain only those speakers with five or more utterances in order to uphold diversity within each speaker's data. Audio samples surpassing the aforementioned scores will be considered into the clean subset, VoxBlink-clean. Through randomly sampling observations, there are very few instances of noisy labels in this subset.

\vspace{-0.5em}
\section{Speaker Verification Model Training}

\begin{table*}[htbp]\centering \footnotesize
\caption{\label{tab::BaseResult} {\it The experimental results of different backbones with/without the VoxBlink-clean dataset. All the benchmarks are based on the cosine scores between trials. \textbf{No post-processing} operations have been employed, such as LMFT, score norm, and QMF. $\Delta$ represents the relative EER reduction on the VoxCeleb1-O trials when using the VoxBlink-clean for Mix-FT compared to not using the VoxBlink-clean.}}
\vspace{+0.5em}
\renewcommand{\arraystretch}{1.1}
\begin{tabular}{@{}llcccccccccc@{}}
\toprule
\multirow{2}*{\textbf{ID}} & \multirow{2}*{\textbf{Model}} & \multirow{2}*{\textbf{Size}} & \multirow{2}*{\textbf{VoxCeleb2}} & \textbf{VoxBlink} & \multirow{2}*{\textbf{$\Delta$}} & \multicolumn{2}{c}{\textbf{VoxCeleb1-O}} & \multicolumn{2}{c}{\textbf{VoxCeleb1-E}}  & \multicolumn{2}{c}{\textbf{VoxCeleb1-H}} \\
\cmidrule(lr){7-12} & ~ & ~ & ~ &  \textbf{-clean} & ~ & \textbf{EER[\%]} & \textbf{mDCF$_{0.01}$} & \textbf{EER[\%]} & \textbf{mDCF$_{0.01}$} & \textbf{EER[\%]} & \textbf{mDCF$_{0.01}$}  \\
\midrule
\multirow{3}*{\textbf{M1}} & \multirow{3}*{ResNet34-TSP} & \multirow{3}*{23.9M} & \texttimes & \checkmark  & - & 2.499 & 0.241 & - & - & - & -  \\
~ & ~ &  & \checkmark & \texttimes & -  & 0.856 & 0.084 & 0.995 & 0.112 & 1.832 & 0.179  \\
~ & ~  &  & \checkmark & \checkmark  & \textbf{13.1\%} & \textbf{0.744} & \textbf{0.057} & 0.988 & 0.109 & 1.787 & 0.176  \\
\midrule
\multirow{2}*{\textbf{M2}} &  \multirow{2}*{ECAPA-TDNN} & \multirow{2}*{14.7M} & \checkmark & \texttimes   & - & 0.856 & 0.081 & 1.078 & 0.118 & 2.059 & 0.197 \\
~ & ~ &  & \checkmark & \checkmark & \textbf{12.5\%} & \textbf{0.749} & \textbf{0.077} & 0.953 & 0.105 & 1.823 & 0.177  \\
\midrule
\multirow{2}*{\textbf{M3}} & \multirow{2}*{SimAM-ResNet100-ASP} & \multirow{2}*{50.2M} & \checkmark & \texttimes  & - & 0.622 & 0.058 & 0.761 & 0.083 & 1.391 & 0.132  \\
~ & ~ &  & \checkmark & \checkmark & \textbf{29.1\%} & \textbf{0.441} & \textbf{0.044} & 0.681 & 0.075 & 1.268 & 0.125  \\
\midrule
\multirow{2}*{\textbf{M4}} & \multirow{2}*{fwSE-ResNet100-ASP} & \multirow{2}*{50.6M} & \checkmark & \texttimes & - &  0.580 & 0.057 & 0.775 & 0.083 & 1.438 & 0.141  \\
~ & ~ &  & \checkmark &  \checkmark &  \textbf{22.1\%} & \textbf{0.452} & \textbf{0.038} & 0.709 & 0.079 & 1.277 & 0.128  \\
\midrule

\end{tabular}
\end{table*}
\vspace{-0.5em}

In this section, we describe the experimental settings and implementation details of several speaker verification systems. We suggest a Mix-FineTune(Mix-FT) \cite{farfield_xiaoyi} training strategy to incorporate the VoxBlink into the training set.

\vspace{-0.5em}
\subsection{Model Settings}

\textbf{ResNet-based model.}
ResNet-based speaker verification model achieved success in past years. Therefore, we conducted experiments using the state-of-the-art(SOTA) ResNet models for comparative analysis. Initially, we employed a standard ResNet34 \cite{cai18_odyssey} followed by a temporal statistic pooling layer as our baseline system. Then, to further tap into the latent potential of the data, we employed a larger ResNet model mounted with attention mechanisms – specifically, ResNet100 with Simple Attention Module (SimAM)\cite{9746294} and frequency-wise Squeeze-Excitation (fwSE)
\cite{thienpondt21_interspeech} modules. The attentive statistics pooling (ASP) is employed to capture the importance of different frames.

\textbf{TDNN-based model.}
ECAPA-TDNN \cite{desplanques20_interspeech} is currently the most popular and SOTA TDNN-series model for speaker verification. We conducted experiments using ECAPA-TDNN with 1024 channels to observe the performance of TDNN on a larger dataset.

\subsection{Implement details}

\textbf{Data Usage.}  We conducted experiments using the VoxCeleb2 development set and the VoxBlink-clean set. The acoustic features are 80-dimensional log Mel-filterbank energies with a frame length of 25ms and a hop size of 10ms. The input frame length is fixed at 200 frames.
 
\textbf{Data Augmentation.}  We adopt on-the-fly data augmentation\cite{cai2020fly} to add additive background noise or convolutional reverberation noise for the time-domain waveform. Also, we apply speaker augmentation with speed perturbation \cite{wang2020dku}. We speed up or down each utterance by a factor of 0.9 or 1.1, yielding shifted pitch utterances that are considered from new speakers. Finally, the training data contains 6,360,345 utterances (3,084,318 from the VoxBlink-clean and 3,276,027 from the VoxCeleb2) from 73,125 speakers (55,143 from the VoxBlink-clean and 17,982 from the VoxCeleb2).
 
\textbf{Training Strategy.} Inspired by \cite{farfield_xiaoyi}, we observe that initiating training both datasets (VoxCeleb2 and VoxBlink-clean) from the start yields similar performance outcomes as beginning solely with the VoxCeleb2 and later incorporating the VoxBlink-clean for fine-tuning. Therefore, we carry most of our experiments based on three stages: 

\textbf{Stage.1 Warm-up.}  We perform a linear warm-up learning rate schedule at the first five epochs to the initial learning rate at 0.1, which aims to prevent model vibration and speed model training.

\textbf{Stage.2 Plateau.} The SGD optimizer updates the model parameters, and the StepLR scheduler with 0.1 initial LR drops to 1e-4 in 30 epochs. The step size is set to 10.

\textbf{Stage.3 Mix-FT.} As the first two stages only use the VoxCeleb2 dev set, we introduce the VoxBlink-clean in the last phase for the Mix-FineTuning(Mix-FT). The training process resumes at 1e-3 LR and gradually drops till convergence.

Finally, we adopt the Equal Error Rate (EER) and Minimum Detection Cost Function (minDCF) to measure system performance. Cosine similarity scores are calculated in the evaluation phase. As for the back end, we utilize the AAM-Softmax \cite{8953658} (m=0.2, s=32) to classify different speakers.
\vspace{-0.5em}
\section{Experimential Results}
\vspace{-0.5em}
\subsection{Base Results}

As shown in table \ref{tab::BaseResult}, the domain of the VoxBlink dataset does not closely align with that of the VoxCeleb2 dataset, as the results on the VoxCeleb1-O trials exhibit better performance when trained only on the VoxCeleb2. However, the VoxBlink-clean can be regarded as a supplementary training set of the VoxCeleb2. As we can see, across models M1 to M4, we achieve performance enhancements of relative 13.1\%, 12.5\%, 29.1\%, and 22.1\% when introducing the VoxBlink-clean, respectively. Performance improvements are observed across all other test protocols (VoxCeleb1-E and VoxCeleb1-H) as well, relatively ranging from 2\% to 12\%. Moreover, as we enlarge the model size, the positive impact of adding the VoxBlink-clean for training becomes increasingly noticeable. 

\subsection{LMFT and Score calibration}
The ASV systems could benefit from several post-processing methods, including Large-Margin Fine-Tune (LMFT) \cite{9414600}, Adaptive Symmetric Score Normalization (AS-Norm) \cite{asnorm} and Quality Measure Functions (QMF) \cite{9414600}. Therefore, we follow the same post-processing settings as \cite{li2023dku} to enhance performance. As shown in Table \ref{tab::Post-Processing}, by incorporating the VoxBlink-clean set for Mit-FT training, followed by a series of post-processing steps, we achieved a reduction in EER from 0.441\% to 0.282\% on the Vox-O test set. Compared to using only the VoxCeleb2 as the training set with post-processing, we achieve a 20.8\% relative EER reduction (0.356\% to 0.282\%).

Nevertheless, without incorporating the VoxBlink-clean for training, the LMFT achieves a 13.7\% improvement, while the Mix-FT using the VoxBlink-clean shows only a 6.1\% boost. We also observed that the EER reduction of the LMFT under the Mix-FT is not that significant in other models. We speculate that this might be because the average speech duration of the VoxBlink is shorter than that of the VoxCeleb2, meaning that less information is carried on for each utterance.

\vspace{-0.5em}

\begin{table}[ht]\centering \footnotesize
    \caption{\label{tab::Post-Processing} {\it The post-processing results based on the SimAM-ResNet100 single system with/without the VoxBlink-clean data in training. }}
    \begin{tabular}{@{}llcccc@{}}
    \toprule
    \multirow{2}*{\textbf{ID}} &\multirow{2}*{\textbf{Method}} & \multirow{2}*{\textbf{$\Delta$}} & \multicolumn{2}{c}{\textbf{VoxCeleb1-O}} \\
    \cmidrule(lr){4-5} & ~ & ~ & \textbf{EER[\%]} & \textbf{mDCF$_{0.01}$}  \\
    \midrule 
    \multicolumn{5}{c}{\textbf{Only VoxCeleb2 for training}} \\
    \midrule

    \textbf{M3} & SimAM-ResNet100 & - & 0.622 & 0.058  \\
    & +LMFT & 13.7\% & 0.537 & 0.045  \\
    & ++ AS-Norm & 8.9\% & 0.489 & 0.047  \\
    & +++ QMF & 27.2\% & 0.356 & 0.040  \\
    \midrule
    \multicolumn{5}{c}{\textbf{VoxCeleb2 for training and VoxBlink-clean for Mix-FT}} \\
    \midrule
    \textbf{M3} & SimAM-ResNet100 & - & 0.441 & 0.044  \\
    & +LMFT & 6.1\% & 0.414 & 0.035  \\
    & ++ AS-Norm & 14.0\% & 0.356 & 0.037  \\
    & +++ QMF & 21.8\% & \textbf{0.282} & 0.029  \\

    \bottomrule
    \end{tabular}
\end{table}

\vspace{-2em}
\section{Conclusion}
This paper introduces a large-scale audio-visual dataset named VoxBlink for the speaker verification task. We develop an automatic multi-modal data-mining pipeline to extract target users' audio-visual segments on YouTube and further conduct multi-modal detectors to build the VoxBlink-clean subset. We also achieve significant improvements by incorporating the VoxBlink-clean into model training across different backbones, which proves that the VoxBlink-clean is an excellent supplementary dataset for training speaker verification models.


\vfill\pagebreak

\newpage
\bibliographystyle{IEEEtran}
{
\fontsize{9pt}{10pt}\selectfont
\bibliography{main}
}

\end{document}